\documentclass[submission, Proceedings]{SciPost}

\hypersetup{
    colorlinks,
    linkcolor={red!50!black},
    citecolor={blue!50!black},
    urlcolor={blue!80!black}
}

\usepackage[bitstream-charter]{mathdesign}
\DeclareSymbolFont{usualmathcal}{OMS}{cmsy}{m}{n}
\DeclareSymbolFontAlphabet{\mathcal}{usualmathcal}

\usepackage{siunitx}
\usepackage{subcaption}

\renewcommand{\eqref}[1]{(\ref{#1})}

\newcommand{\figref}[1]{Figure~\ref{#1}}

\newcommand{\secref}[1]{Section~\ref{#1}}

\numberwithin{equation}{section}
\numberwithin{figure}{section}
\numberwithin{table}{section}

\begin{document}



\begin{center}{\Large \textbf{
The \emph{muX} project \\
}}\end{center}

\begin{center}
F.~Wauters\textsuperscript{1$\star$} and
A.~Knecht\textsuperscript{2} on behalf of the \emph{muX}
collaboration\footnote{https://www.psi.ch/en/ltp/mux} 
\end{center}

\begin{center}
{\bf 1} PRISMA+ Cluster  of  Excellence  and  Institute  of  Nuclear
Physics, Johannes  Gutenberg  Universit\"at  Mainz,  Germany 
\\
{\bf 2} Paul  Scherrer  Institut,  Villigen,  Switzerland
 \\
* Corresponding: fwauters@uni-mainz.de
\end{center}

\begin{center}
\today
\end{center}

\definecolor{palegray}{gray}{0.95}
\begin{center}
\colorbox{palegray}{
  \begin{tabular}{rr}
  \begin{minipage}{0.05\textwidth}
    \includegraphics[width=24mm]{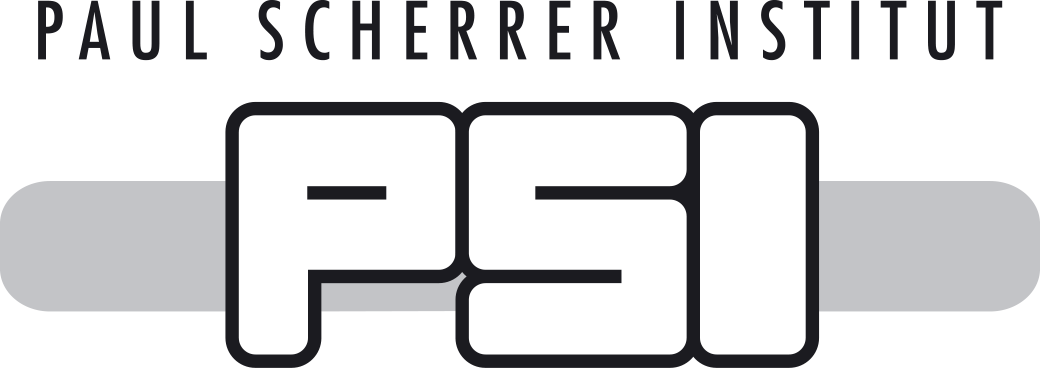}
  \end{minipage}
  &
  \begin{minipage}{0.82\textwidth}
    \begin{center}
    {\it Review of Particle Physics at PSI}\\
    \doi{10.21468/SciPostPhysProc.2}\\
    \end{center}
  \end{minipage}
\end{tabular}
}
\end{center}

\section*{Abstract}
{\bf \boldmath The \emph{muX} project is conducting a series of muonic
  X-ray measurements in medium- and high-Z nuclei at PSI, utilizing a high-purity germanium detector array, in-beam
  muon detectors, and a modern digital data-acquisition system. A
  novel hydrogen target for muon transfer was developed, enabling measurements
  with as little as a few micrograms of target material.  First
  measurements with radioactive Cm and Ra targets were conducted,
  aimed at determining their nuclear charge radii.  These serve as
  important input for upcoming atomic parity violation
  experiments. The apparatus is also used to perform a feasibility
  study of an atomic parity violation experiment with the $2s-1s$
  muonic X-ray transition. In addition, the setup has been made
  available for a wider range of nuclear, particle, and solid-state
  physics measurements.}

\setcounter{section}{22}
\label{sec:muX}

\subsection{Introduction}
\label{muX:sec:intro}

Muonic atoms are exotic atoms that form when negative muons are
stopped in a target and are subsequently captured by a nearby atom in
a highly excited atomic orbital of $n\geqslant$14. The muons quickly
cascade down to the $1s$ orbital, initially predominantly via Auger
transitions: at lower $n$ radiative transitions take over. As the
muon mass is about 207 times larger than the electron mass, the
muonic X-rays range in energy from a few tens of keV for low-Z nuclei
to several MeV for heavier nuclei.  The capture and cascade processes
occur on (sub)nanosecond timescales.  The emitted radiation therefore
appears prompt relative to a muon stopping in the target. Once in
the $1s$ orbit, the muon either decays in orbit, or is captured
by the nucleus. The latter is the dominant decay channel for Z=12
and above ~\cite{PhysRevC.35.2212}.

Muonic atoms have proven to be a valuable tool to measure nuclear
properties and probe short-range interactions between the muon and the
nucleus. With the Bohr radius of the muon compared to the electron
scaling as $m_e/m_{\mu}$, there is substantial overlap between
the muon and nuclear wave functions. Finite size effects are thus
highly amplified. In the past, the absolute nuclear charge radii
$<r^2>^{1/2}$ of almost all stable nuclei have been determined with a
typical accuracy of $10^{-4}$~-~$10^{-3}$ by measuring the $2p-1s$
transition energy~\cite{FRICKE1995177}. More recently, the radii of
the lightest nuclei were measured by the CREMA collaboration (Section~21~\cite{section21}) using
laser spectroscopy on muonic
atoms~\cite{Pohl:2016, Antognini:1900ns, Pohl:2016xsr, Krauth:2020}.

Formerly, this approach was limited to stable isotopes, as a sufficient
amount of target material is needed to stop a $\mu^-$ beam with a
momentum of typically 30~MeV/$c$. This excludes many interesting nuclei,
such as the highly-deformed radium isotopes. Radium is a prime
candidate for an Atomic Parity Violation (APV) experiment, using laser
spectroscopy on a trapped ion~\cite{Portela:2013twa,Fan:2019eze},
where the Parity Non-Conserving (PNC)  $E1_{PNC}$ atomic $S-D$ transition is
proportional to $K_r Z^2 Q_W$, with $Q_W$ the weak nuclear charge, and
$K_r$ a relativistic enhancement factor which depends on the nuclear
charge radius~\cite{PhysRevA.78.050501}. The \emph{muX} collaboration
aims to determine this radius by measuring the $2p-1s$ transition
energy of $^{226}$Ra ($T_{1/2}$=1600~y.). For this we have developed a
novel technique, stopping muons in a high-pressure H$_2$/D$_2$ target,
using a sequence of transfer reactions to efficiently stop
muons in a few micrograms of target material. This technique was first
established with gold targets, then applied to $^{226}$Ra and
$^{248}$Cm (see \secref{muX:sec:radioactive}).

With fundamental interactions being our primary physics motivation,
the collaboration is also investigating the possibility of measuring APV
directly in muonic atoms. A neutral parity-violating interaction mixes
the $2s_{1/2}$ and $2p_{1/2}$ atomic levels, resulting in an E1
admixture in the otherwise pure M1 $2s_{1/2}-1s_{1/2}$
transition. Measuring such a parity-odd observable was first reviewed
by Feinberg \& Chen~\cite{Feinberg1974} and Missimer \&
Simons~\cite{Missimer:1984hx}. More recently, the possibility of searching 
for interactions between the muon and the nucleus beyond the Standard Model led to revived
interest~\cite{Batell2011,McKeen2012}. While the PNC effect is 
largest for low-Z atoms, separating the radiative M1/E1
transition from other transitions in the cascade severely complicates
the design of such an experiment~\cite{PhysRevLett.78.4363}. We focus
on Z$\simeq$30 nuclei, where the single-photon $2s-1s$ transition
becomes the dominant path depopulating the $2s$ level. The current
goal of the collaboration is to isolate the transition in the cascade,
and to significantly improve the signal-to-background ratio in the
region-of-interest (ROI) in the X-ray spectrum~(see
\secref{muX:sec:2s1s}).

Since 2015 we have been developing an advanced muonic X-ray experimental
setup, combining a high-purity germanium (HPGe) detector array and a
modern data-acquisition system (DAQ) with various target
configurations. The setup is currently also being used for
non-destructive elemental analysis, muon-capture studies probing
matrix elements of interest for neutrinoless double $\beta$ decay, and
further nuclear-charge radius measurements of various radioactive
elements and rare isotopes.


\subsection{Experimental setup.}
\label{muX:sec:setup}

\begin{figure}[!ht]
\begin{minipage}[t]{0.5\linewidth}
\centering
\includegraphics[width=\textwidth]{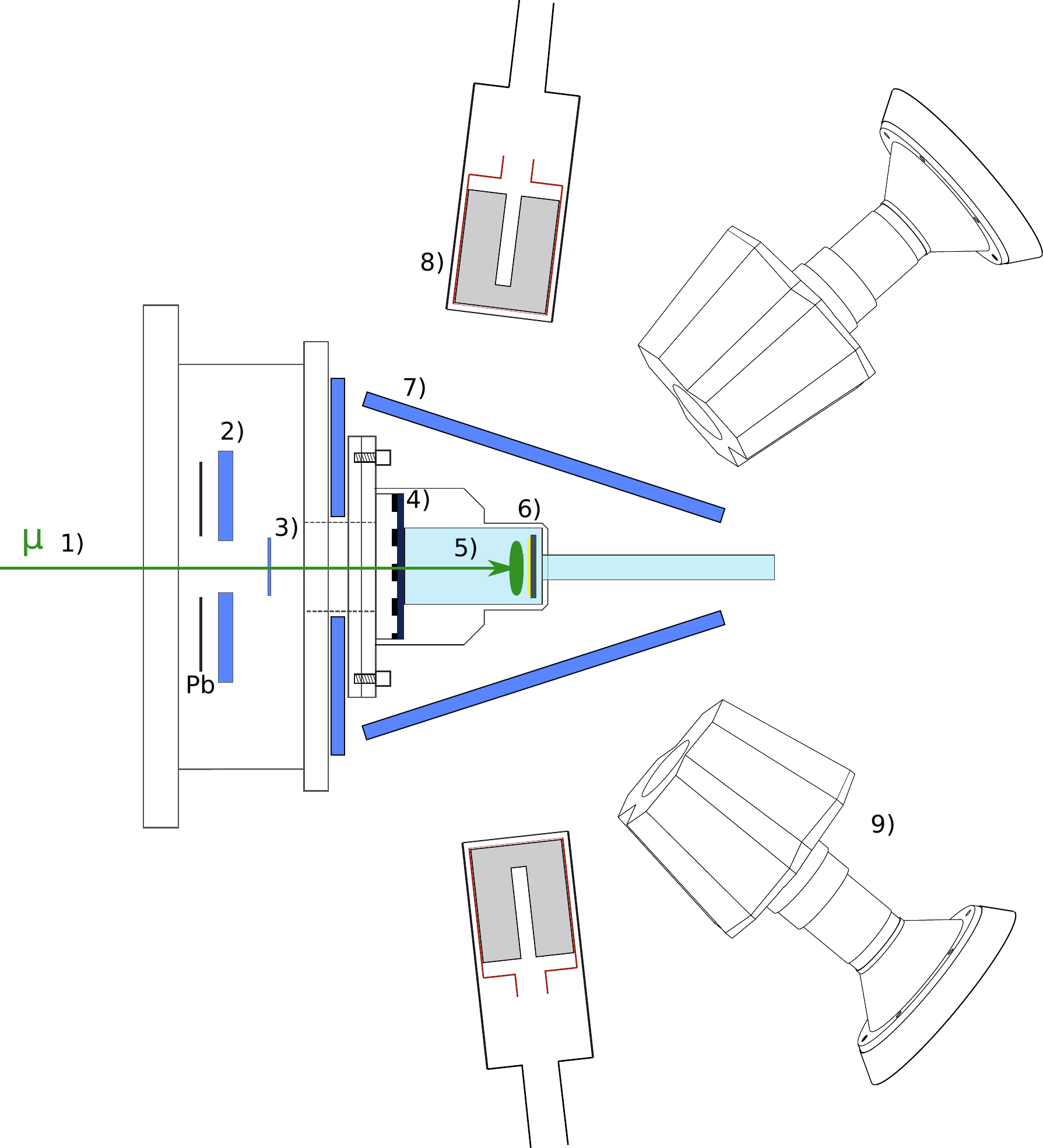}\captionsetup{width=.9\linewidth}
\caption{The \emph{muX} setup, with 1)~the $\mu^-$ beam passing through 2)~a
  veto detector with a 18~mm aperture, and 3)~a 200~$\mu$m thick muon
  detector. The cell 4) with ~a 600~$\mu$m carbon fibre window
  supported by a Ti grid holds 5)~100~bar of hydrogen gas, with the 6)
  target mounted in the back. 7)~Electron veto detectors. 8)~Standard
  and 9)~MiniBall cluster HPGe detectors.}
\label{fig:muX:Setup}
\end{minipage}
\hfill
\begin{minipage}[t]{0.45\linewidth}
\centering
\includegraphics[width=\textwidth]{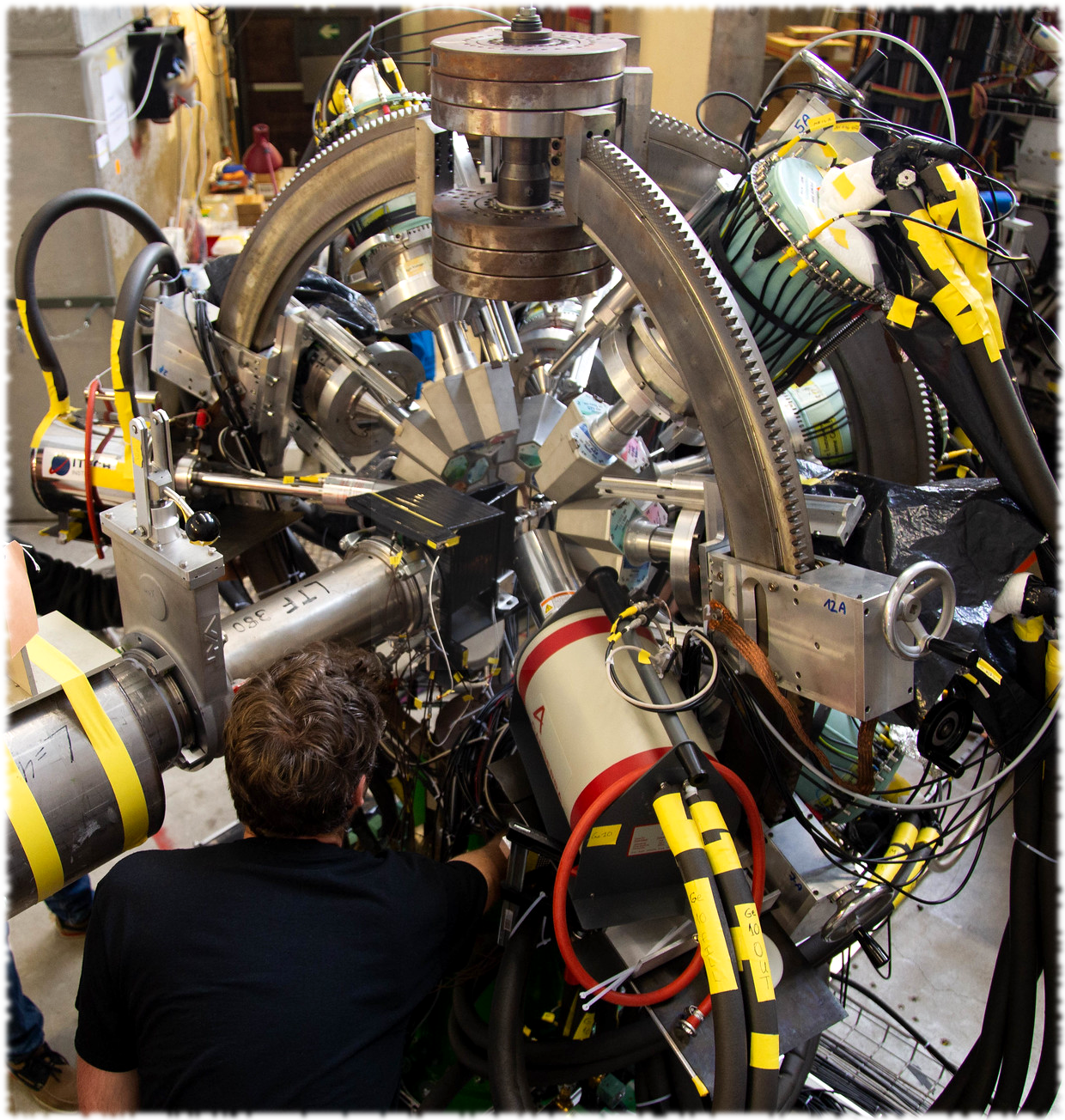}\captionsetup{width=.9\linewidth}
\caption{The MiniBall array with eight cluster detectors complemented
  by a 70~\% coaxial detector and a low-energy planar detector
  installed at the $\pi$E1 beamline for the 2019 experimental run,
  with the \emph{muX} beam snout. The target cell is covered by the black
  electron detectors.}
\label{fig:muX:detector_array}
\end{minipage}
\end{figure}

The \emph{muX}
apparatus~(\figref{fig:muX:Setup}~and~\figref{fig:muX:detector_array})
is located at the $\pi$E1 beam-line of PSI, where a typically
30-40~MeV/c $\mu^-$ beam with a momentum width $\Delta p/p$ of 3~\% passes through an electron separator before reaching 
the experiment. A custom beam snout houses an in-vacuum
set of beam counters, thin plastic scintillator slabs read out by
SiPMs, a lead target mounted away from the beam axis for calibration
purposes, and a port for directly mounting various targets, thereby
minimizing scattering of the low-energy muons.

The target itself is surrounded by 5~mm thick plastic scintillators,
efficiently detecting outgoing decay electrons, thus enabling
various cuts on the data such as suppressing Brems\-strah\-lung background
in the HPGe detectors.

The \emph{muX} HPGe detector array is constructed from various
detectors provided by the collaborating institutions. Early campaigns,
such as the $^{185/187}$Re measurement aimed at determining the charge
radii and quadruple
moments~\cite{PhysRevC.101.054313}, were
conducted with just a few coaxial HPGe detectors. For the 2017 and
2018 campaigns, 7 compact coaxial detectors from the French/UK loan
pool\footnote{https://gepool.in2p3.fr/} with relative efficiencies of
around 60\% and one Miniball cluster detector were added. In the
summer of 2019, the full MiniBall detector array~\cite{Warr:2013ana}
was installed at the $\pi$E1
beamline~(\figref{fig:muX:detector_array}), operating for a 7 week
measurement campaign. The \emph{muX} automatic liquid-nitrogen filling
system enables extended continuous operation of the HPGe detectors.

The MIDAS-based DAQ uses SIS3316 250 MSPS
digitizers\footnote{https://www.struck.de/sis3316.html} which record
all detector hits above threshold. Physics events are reconstructed
offline by the analysis software. A digital filter running on the
digitizer module FPGA integrates the detector signals, in addition, a
section of the raw waveform is saved for offline analysis, where a
time resolution of better than 10~ns~(FWHM) for the HPGe detector hits
is achieved.

\subsection{Radioactive target measurements}
\label{muX:sec:radioactive}

\begin{figure}[!ht]
\begin{minipage}[b]{0.52\linewidth}
\centering
\includegraphics[width=\textwidth]{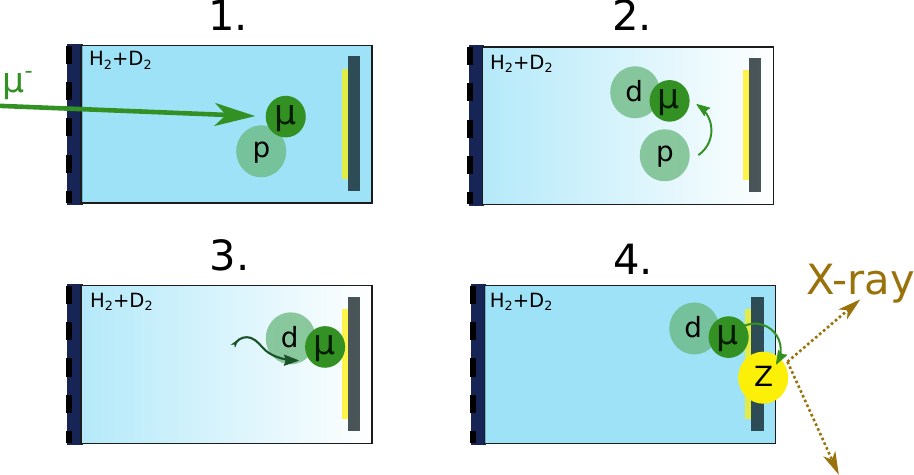}\captionsetup{width=.9\linewidth}
\caption{1. After slowing down a $\mu$p atom is formed. 2. In
  $\mathcal{O}$(100)~ns, the muon transfers to deuterium, gaining
  45~eV in kinetic energy. 3. After scattering down in energy to
  around 4~eV, the $\mu$d-H$_2$ scattering cross section becomes
  negligibly small, and the $\mu$d atom travels straight until it hits
  a wall or our target, where 4. the $\mu^-$ transfers to a high-Z
  atom.}
\label{muX:fig:transfer}
\end{minipage}
\hspace{0.2cm}
\begin{minipage}[b]{0.48\linewidth}
\centering
\includegraphics[width=\textwidth]{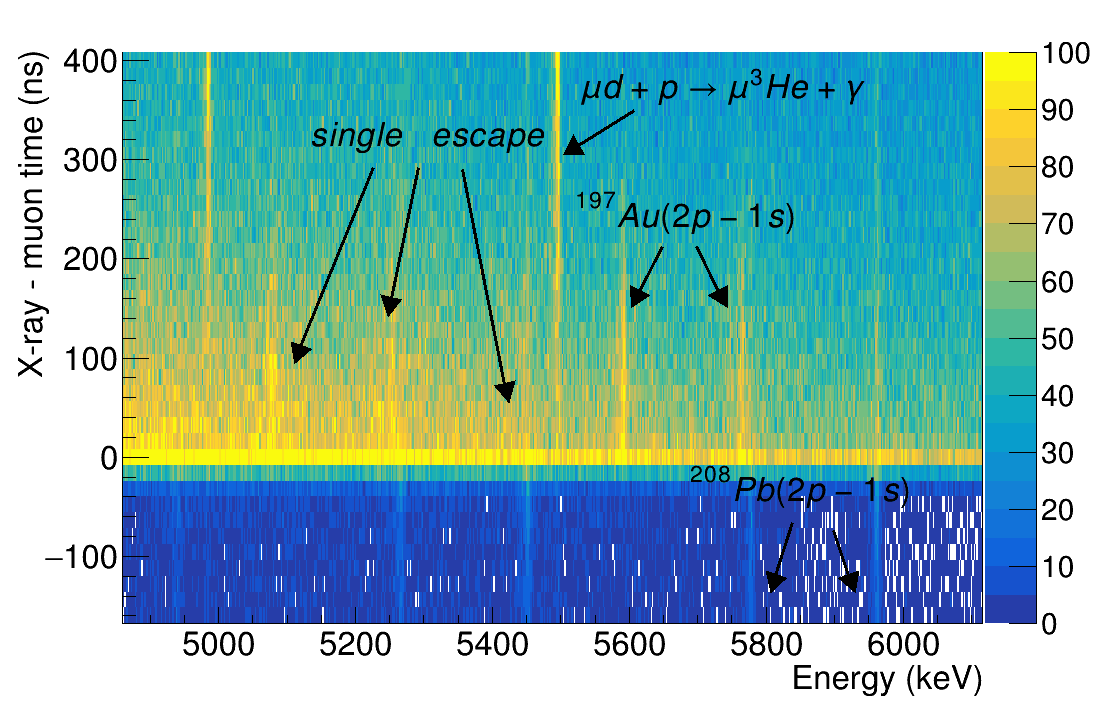}\captionsetup{width=.9\linewidth}
\caption{Muonic X-ray energies versus their time relative to an
  incoming muon. X-rays from direct stops appear at 0~ns. The Au
  X-rays appear over $\mathcal{O}$(100)~ns, the typical timescale for
  the transfer processes. The background mainly consists of decay
  electrons, and neutrons emitted after nuclear muon capture.}
\label{muX:fig:gold}
\end{minipage}
\end{figure}

One of the principal goals of the \emph{muX}
project~\footnote{Proposal R-16-01} is to measure the $2p-1s$
transition energies for $^{226}$Ra, a radioactive isotope for which
the maximum allowed quantity in the experimental area is 5~$\mu$g. As
the stopping power of such a low-mass target is insufficient by orders
of magnitude, the \emph{muX} collaboration has developed a novel
method, stopping muons in a small 100~bar H$_2$ target with a small
admixture of D$_2$. Through a series of transfer reactions the muon is
transported to the target material mounted at the back of the
cell~(\figref{muX:fig:transfer}), hereby exploiting the
Ramsauer-Townsend effect~\cite{PhysRevA.73.034501,Adamczak:2006tjg,
  Adamczak:2006xvz, Strasser:2009ttm}, which causes H$_2$ gas to
become almost fully transparent for a $\mu$d atom.

After a first optimization of the target geometry and conditions with
Monte-Carlo simulations, the transfer method was established by
mounting a thin gold target at the back of the cylindrical gas
cell. The beam momentum and deuterium concentration were optimized for
the number of gold X-rays per muon, after which a small 3~nm thick
gold target was installed. A total stopping efficiency per beam muon
of 1.2~\% was achieved for this 5~$\mu$g target~(see
\figref{muX:fig:gold}).

In order to have an efficient transfer target, it is imperative that
the (radioactive) material is deposited as a uniform surface
layer. Due to the low kinetic energy of the $\mu$d atom, an organic
surface layer of $>$100~nm acts as a barrier and significantly reduces
the transfer efficiency, rendering traditional molecular plating
techniques inadequate.  Several $^{248}$Cm and $^{226}$Ra targets were
produced at the Institute of Nuclear Chemistry of the Johannes
Gutenberg University Mainz, combining a custom electro-deposition
technique combined with a novel \emph{drop-on-demand} method where
micro-drops of activity in solution are deposited on glassy carbon
disks, the low-Z backing material of the target~\cite{HAAS201743}.

\begin{figure}[!ht]
\centering
\includegraphics[width=0.75\textwidth]{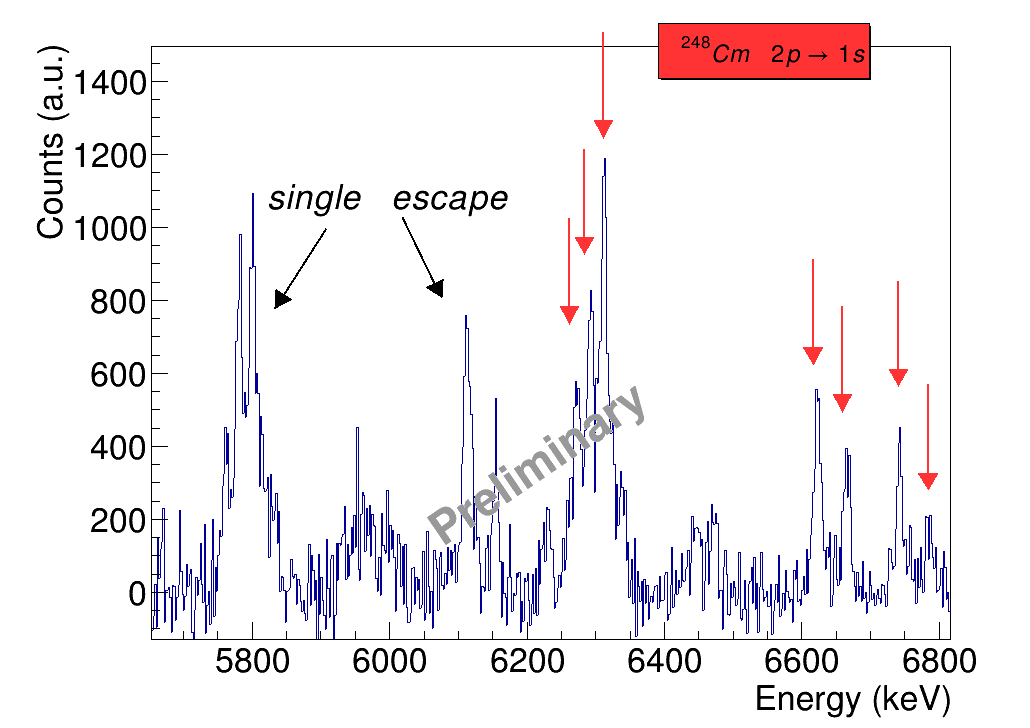}\captionsetup{width=.9\linewidth}
\caption{The $^{248}$Cm muonic X-ray spectrum from the hydrogen
  transfer cell after subtracting the lead calibration lines and the
  $\gamma$ background from muon capture on $^{16}$O.}
\label{muX:fig:Cm}
\end{figure}

\figref{muX:fig:Cm} shows the muonic X-rays from $^{248}$Cm measured
during the 2019 campaign with a 15~$\mu$g curium target. After
subtracting several background contributions, the $2p-1s$ transitions
are clearly visible. Despite having nuclear ground state of spin 0,
the energy scale of high-Z muonic atoms is such that the muon spin
couples to excited nuclear states with a non-zero
spin~\cite{PhysRevC.1.1184,PhysRevA.99.042501}. This leads to a
complicated dynamic hyperfine structure in the observed transition
energies, which needs to be understood to extract the nuclear charge
radius from the data. The largest uncertainty in the calculations of
the transition energies is caused by the two-photon exchange nuclear
polarization~\cite{Bergem:1988zz,Haga:2002zz}.

In addition to the $^{248}$Cm target, two $^{226}$ Ra targets were
used. The data obtained are currently under analysis to determine
whether the X-ray yield is sufficient to achieve the necessary
accuracy on the nuclear charge radius.

\subsection{Extended experimental program}
\label{muX:sec:program}
\subsubsection{2s-1s measurements}
\label{muX:sec:2s1s}

With an expected branching ratio of $\mathcal{O}$(10$^{-4}$) for the
single-photon $2s-1s$ muonic X-ray transition in the cascade of
Z$\simeq$30 atoms, a possible APV experiment with a PNC observable
using this transition is severely hampered by an overwhelming
background in the energy region of interest (ROI) from scattered
$(n\geqslant 3)p-1s$ X-rays, Bremsstrahlung from decay electrons, and
neutrons from muon capture. For this reason this transition has never
been observed. The goal of the \emph{muX} project is to observe this
transition, significantly improve the signal-to-background in the ROI,
and determine the reach of a possible APV experiment.

\begin{figure}[!ht]
\begin{minipage}[b]{0.55\linewidth}
\centering
\includegraphics[width=\textwidth]{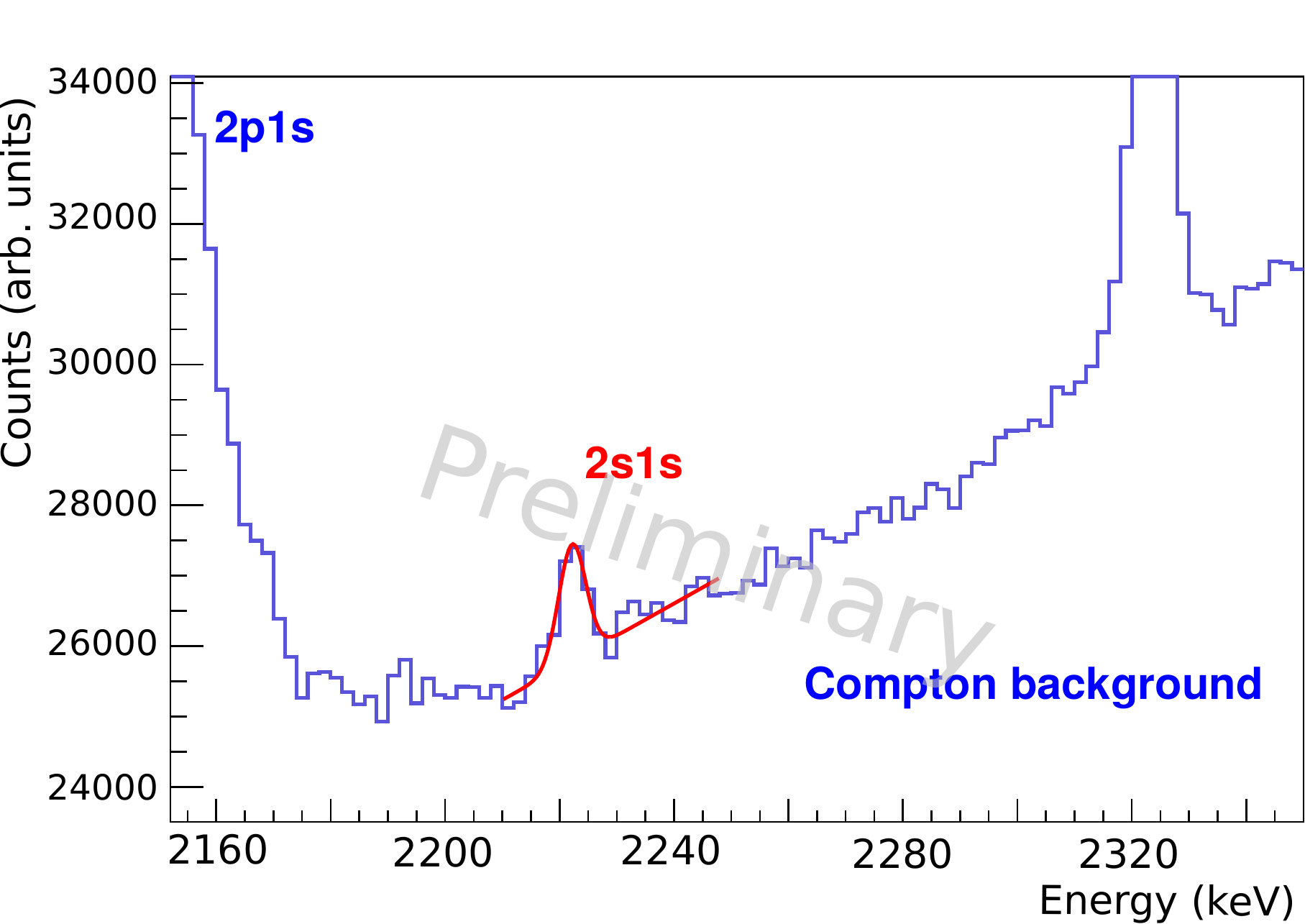}\captionsetup{width=.9\linewidth}
\caption{The $2s-1s$ full energy peak of muonic Kr clearly visible at 2.22~MeV
  above the Compton background of $(n>2)p-1s$ transitions after
  subtracting background $\gamma$'s from nuclear muon capture
  processes.}
\label{muX:fig:Kr}
\end{minipage}
\hspace{0.2cm}
\begin{minipage}[b]{0.45\linewidth}
\centering
\includegraphics[width=\textwidth]{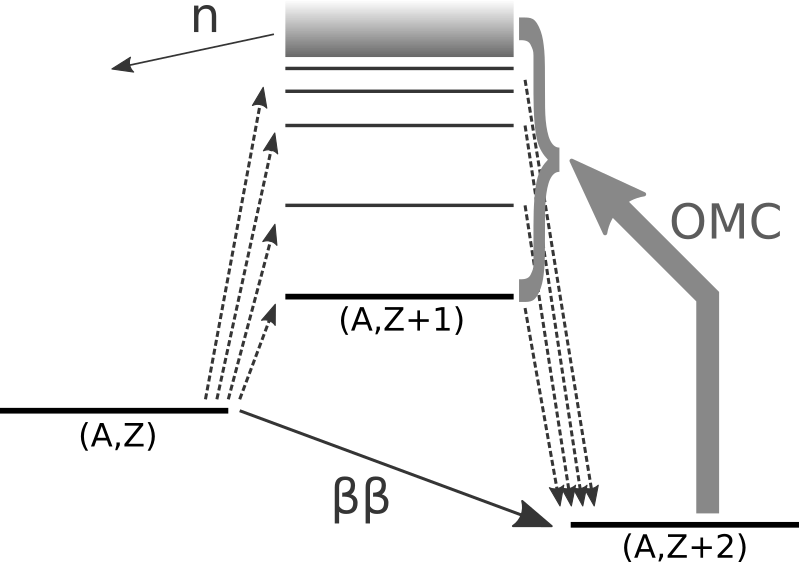}\captionsetup{width=.9\linewidth}
\caption{Partial muon capture rates of selected isotopes provide
  access to the transition strengths via virtual states in double
  $\beta$-decays.}
\label{muX:fig:dBeta}
\end{minipage}
\end{figure}

The initial average orbital quantum number $l$ after $\mu H
\rightarrow \mu Z$ transfer is lower than the initial $l$
for direct atomic capture~\cite{HAFF1977363}. We have observed that as
a consequence, the $2s$ population in the cascade of Ar, Kr, and Xe
is increased by a factor of 3-4, thus increasing the branching
ratio of the $2s-1s$ transition. A 7~day measurement with a
100~bar H$_2$ target and an 0.1~\% Kr addition was performed. After
subtracting the nuclear capture background from muon stops in the
surrounding materials, the $2s-1s$ full energy peak is clearly
visible, achieving a signal to background of about 1/10
(\figref{muX:fig:Kr}).

To further reduce the background in the ROI, the transitions feeding
the $2s$ level were used to tag events of interest. While sacrificing
efficiency, this approach significantly reduces the
background: the continuous Compton background from e.g. $3p-1s$
photons is fully eliminated, and the accidental background from
neutrons and decay electrons is at the same level as the signal yield,
which can be further reduced by improving the time resolution. The
only remaining challenging background is the satellite peaks introduced in the spectra by Compton scattered photons with energy depositions in
the region of the $2s$~feeding transitions. This background needs to
be controlled by optimizing the detector geometry. During the 2019
campaign, one week of data was taken with such an optimized geometry,
collecting over 10$^{11}$ muon stops on an isotopically pure $^{64}$Zn
target.


\subsubsection{Other measurements}

To fully benefit from the availability of the MiniBall detector array,
the \emph{muX} experimental program was expanded in 2019. The partial
ordinary muon capture rates on enriched $^{130}$Xe, $^{82}$Kr,
and $^{24}$Mg to specific excited states in the daughter nucleus were
measured. Such measurements provide valuable information to determine
the nuclear matrix elements in neutrinoless double
$\beta$-decay~\cite{PhysRevC.99.024327, Jokiniemi:2020ydy}, as these
states act as intermediate virtual states in the double $\beta$-decay (\figref{muX:fig:dBeta})
of isotopes such as $^{130}$Te~\cite{PhysRevLett.124.122501,
  Andringa:2015tza} and $^{82}$Se~\cite{Arnold:2010tu}.

In addition, the \emph{muX} apparatus was made available to perform
elemental analysis on a series of cultural heritage samples, 17th
century Japanese coins and an ancient Chinese mirror, significantly
improving the sensitivity of previous J-PARC
measurements~\cite{Ninomiya:2015ata}, and a number of coins and
recently found artifacts from the Roman Augusta Raurica site, nearby
PSI. The intense muon beam and efficient detector setup permitted a
narrowly collimated beam, probing different areas of a sample. Muonic
X-ray spectroscopy provides information about the bulk material
compared to the surface sensitivity of traditional fluorescence X-ray
analysis. Furthermore, for high Z-elements such as lead the isotopic
composition can be extracted.

\subsection{Conclusions and Outlook}
\label{muX:sec:outlook}
The \emph{muX} efforts have resulted in a revived muonic X-ray program
at the Paul Scherrer Institut. A new versatile experimental setup
allows us to efficiently take data for extended periods of time.

The new hydrogen transfer target we have developed enables muonic
X-ray measurements with a very small amount of target material. First
measurements were performed with a Cm and Ra targets, with the purpose
of extracting the nuclear charge radius, providing valuable input for
upcoming APV experiments. The radioactive program will be extended to
other elements, aiming to measure the third of three isotopes of odd
Z-elements needed to calibrate the vast amount of isotope shift data
available from laser spectroscopy on radioactive
elements~\cite{Chael2012}.

The single photon $2s-1s$ transition in the muonic X-ray cascade was
observed for the first time, and significant progress was made in
reducing the backgrounds. This opens up the possibility for an APV
experiment with a sensitivity of $\mathcal{O}$(1) of the Standard
Model amplitude, i.e., such a measurement would act as a new physics
search.

The two additional measurements of the 2019 campaign, the OMC capture
measurements and the elemental analysis, will continue as separate
projects with the support of the \emph{muX} collaboration.

\subsubsection*{Acknowledgements}
This work was supported by the PSI through the
Career Return Program, by the Swiss National Science Foundation
through the Marie Heim-V\"{o}gtlin grant No.~164515 and the project
grant No.~200021165569, by the Cluster of Excellence "Precision
Physics, Fundamental Interactions, and Structure of Matter" (PRISMA
EXC 1098 and PRISMA+ EXC2118/1) funded by the German Research
Foundation (DFG) within the German Excellence Strategy (ProjectID
39083149), and by the DFG under Project WA 4157/1. We wish to thank
all collaborating institutions providing essential instrumentation
such as high-purity germanium detectors.

\bibliography{muX}

\nolinenumbers

\end{document}